# Generation mechanism of cell assembly to store information about hand recognition

Takahiro Homma[1*]

[1] *University of Electro-Communications, 1-5-1 Chofugaoka, Chofu, Tokyo, Japan*

Abstract

A specific memory is stored in a cell assembly that is activated during fear learning in mice; however, research regarding cell assemblies associated with procedural and habit learning processes is lacking. In modeling studies, simulations of the learning process for hand regard, which is a type of procedural learning, resulted in the formation of cell assemblies. However, the mechanisms through which the cell assemblies form and the information stored in these cell assemblies remain unknown. In this paper, the relationship between hand movements and weight changes during the simulated learning process for hand regard was used to elucidate the mechanism through which inhibitory weights are generated, which plays an important role in the formation of cell assemblies. During the early training phase, trial and error attempts to bring the hand into the field of view caused the generation of inhibitory weights, and the cell assemblies self-organized from these inhibitory weights. The information stored in the cell assemblies was estimated by examining the contributions of the cell assemblies outputs to hand movements. During sustained hand regard, the outputs from these cell assemblies moved the hand into the field of view, using hand-related inputs almost exclusively. Therefore, infants are likely able to select the inputs associated with their hand (that is, distinguish between their hand and others), based on the information stored in the cell assembly, and move their hands into the field of view during sustained hand regard.

[*] Email address: takahiro.homma@uec.ac.jp

## 1. Introduction

Humans and animals learn daily and memorize new information acquired by learning. Changes in the brain that correspond to memory are called memory engrams, and the particular population of neurons that hold a memory engram is called a cell assembly. The neurons that compose a cell assembly are connected by strong synaptic connections to each other (Hebb, 1949).

Many studies have attempted to prove the cell assembly hypothesis using several different approaches (Tonegawa, Liu, Ramirez, & Redondo, 2015). The ablation (Han, et al., 2009) or inhibition (Zhou, et al., 2009) of a subset of active neurons during training was shown to disrupt fear memories in mice. Moreover, mice could recall fear memories following the optogenetic reactivation of neurons that were activated during fear learning (Liu, et al., 2012). However, to identify cell assemblies for other types of memories, such as procedural memories, the development of other methods may be required because "procedural or habit memories develop slowly with multiple rounds of training" (Tonegawa, et al., 2015).

During modeling studies, in contrast with experimental studies, the simulated learning of hand regard, which is a type of procedural learning, has demonstrated cell assembly formation (T. Homma, 2018). Hand regard refers to the behavior of repeatedly looking at one's own hands and is often observed in infants from 2 to 3 months of age. After experiencing hand regard, infants may recognize their own hands. A previous study attempted to simulate the recognition of one's hand through hand regard.

Other modeling studies examining the formation of cell assemblies have been conducted. Inhibition during learning is essential to the formation and segregation of cell assemblies (Buzsaki, 2010); therefore, inhibitory interneurons have been incorporated into models in advance (Lansner, 2009; Tomasello, Garagnani, Wennekers, & Pulvermuller, 2017).

However, hand regard occurs during a period of rapid brain development, to approximately 3 months of age, and little is known regarding what part of the brain is associated with the learning of hand regard; therefore, whether inhibitory connections exist in the brain regions associated with the learning of hand regard remains unclear.

For these reasons, in the previous model, the weights of the neural networks were initialized randomly, within the range [-0.1, 0.1] (T. Homma, 2018). Hand regard represents a form of procedural learning, in which the hand is moved to the center of the field of view. The simple neural network was trained using a modified, real-time recurrent learning (RTRL) algorithm (Williams & Zipser, 1989), to incorporate time-varying inputs and outputs during hand regard. During the training phase, units in the hidden layer (hidden units) gradually became interconnected with the inhibitory weights, and most of the weights between the hidden units became inhibitory. Then, a group of hidden units that were strongly coupled by excitatory weights appeared, representing a cell assembly. The hidden units belonging to the cell assembly were the multiple winners of excitatory and inhibitory competitions, which could be understood as a phenomenon based on a principle of soft Winner-Take-All or k-Winner-Take-All (Maass, 2000; Majani, Erlanson, & Abu-Mostafa, 1989).

However, some problems remained unresolved in the previous study. Since the weights of the neural networks were initialized randomly, within the range [-0.1, 0.1], about half of the initial weights are inhibitory. Even the presence of initial inhibitory weights near zero does not lead to the formation of the cell assembly. As shown in section 4.2, the cell assembly is formed according to the principle of the soft Winner-Take-All or k-Winner-Take-All only after most of the weights become inhibitory and those absolute values become larger than the initial values. Therefore, to know the generation mechanism of cell assembly, it is necessary to elucidate the process in which most of the weights between the hidden units became inhibitory during the early training phase.

Moreover, the nature of the information stored in the self-organized cell assembly was not clear. To distinguish the self from the other, predicted sensory feedback and actual sensory feedback were compared (Decety & Sommerville, 2003). The previous study suggested that predicted sensory feedback (corollary discharge) and actual sensory feedback (visual and proprioceptive inputs) were compared, and the network acquired the ability to distinguish between a hand and other objects. What the network could distinguish between the hand and the other objects means that the network could recognize the hand. However, whether the information associated with this ability was stored in the self-organized cell assembly remains unclear. The purpose of the present study is to advance the goals of the previous study by addressing these unresolved questions. The elucidation of this mechanism and the determination of what information is stored in the cell assembly may be applicable to studies examining the cell assemblies associated with other forms of procedural learning.

The RTRL algorithm that was adopted in the previous study can calculate the weight changes associated with motor-command errors. Therefore, the relationship between hand movements and weight changes was examined. In the early training phase, trial and error attempts to bring the hands into the field of view leads to weight reductions and the generation of inhibitory weights. The information stored in the cell assembly was estimated by examining the contributions of the outputs from the cell assemblies to hand movements. From these results, the cell assembly was found to store information necessary to recognize hands and to move them to the center of the field of view. Hand regard disappeared in the previous model, and hand regard disappears in infants at approximately four months of age (White, Castle, & Held, 1964). The disappearance of hand regard from the previous study was determined to be caused by the saturation of the cell assembly.

## 2. Previous study

In the previous study, a learning model for hand regard that self-organizes cell assemblies was constructed, as follows (T. Homma, 2018). For simplicity, the left hand and right hands of the infant and a target object are each denoted by one square in a two-dimensional space (Fig. 1a), and the structure of the upper limbs was omitted from the model; coordinate transformations (which translate sensory inputs to motor outputs) were omitted, and a simulation calculation was executed in a two-dimensional extrinsic coordinate frame. Hereafter, within the model, one hand of the infant, both hands of the infant, an object other than the hands, and more than one object other than the hands are respectively referred to as "hand", "hands", "other" and "others".

The network architecture of the model, which is composed of a three-layer network, is shown in Fig. 1b. The first input layer includes an array of 238 input units, which receive visual inputs, proprioceptive inputs, and corollary discharges. The second hidden layer consists of 48 hidden units, which project to eight output units in the third output layer. Each hidden unit receives inputs from all input units, and each output unit receives inputs from all hidden units. Four of the output units control the movements of the left "hand", and the other four control the movements of the right "hand".

Concerning inputs, self-body recognition in adults can be reduced to two senses of the self; namely, a sense of self-ownership and a sense of self-agency, which are considered to emerge mainly from the integration of visual and proprioceptive/tactile inputs and the integration of these inputs and efference copy, respectively (Jeannerod, 2003; Shimada, Qi, & Hiraki, 2010). Under the supposition that an infant recognizes their own hands through the learning of hand regard, it is natural to conjecture that this learning has some relation with the sense of self-ownership and the sense of self-agency. For this reason, it was hypothesized that inputs of this learning were corollary discharges, visual and proprioceptive inputs, and the network that simulated the areas of the brain related to the sense of self-ownership and the sense of self-agency was adopted. Specifically, the hidden units consisted of two groups. The units in the first group were associated with the sense of self-agency and received and integrated corollary discharges, visual inputs, and proprioceptive inputs from the input units. The units in the second group were associated with the sense of self-ownership and received and integrated visual inputs and proprioceptive inputs from the input units. However, it is not yet clear whether there is a difference between the contributions of the two groups in distinguishing "hand" from "other".

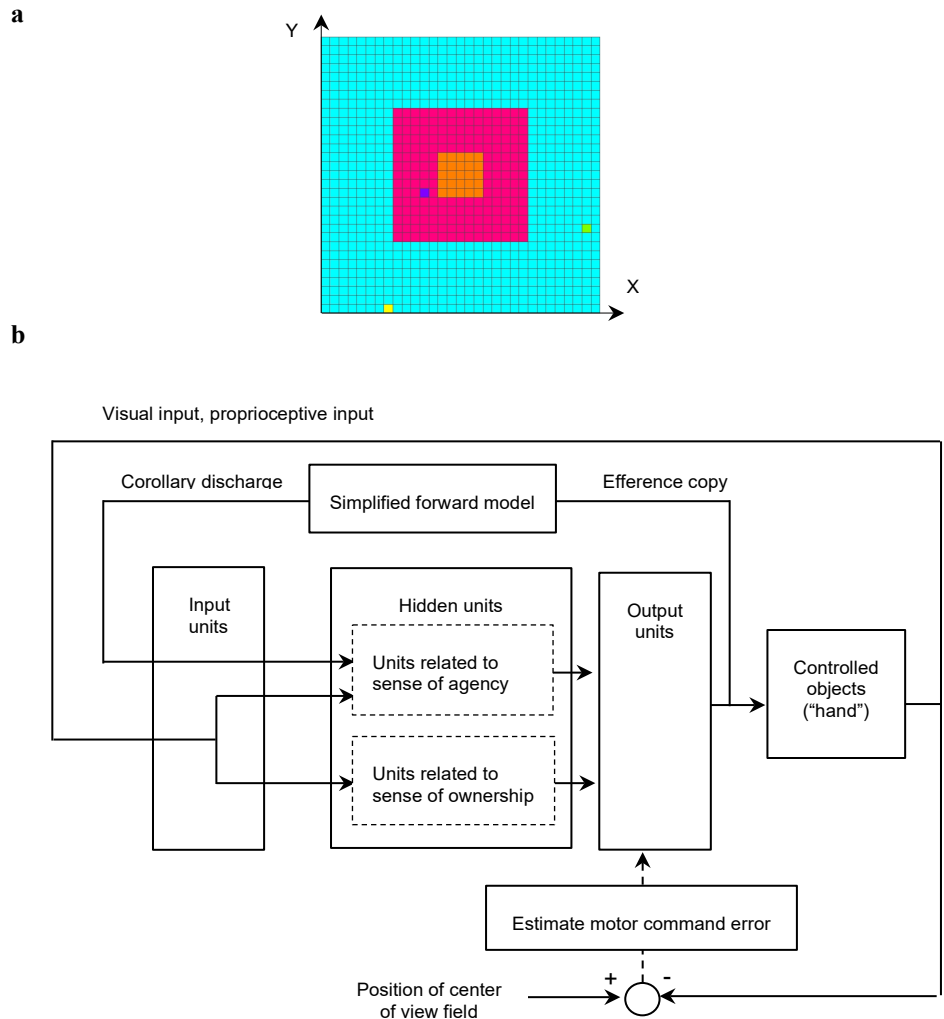

**Fig. 1.** Simulation model for learning hand regard: (a) Infant's field of view and the reachable area of the infant's hands and other objects. The left hand and right hand of the infant and the other object, which are represented by the yellow, yellow-green, and blue squares, respectively, can move to the blue, red, and orange areas. The width corresponds to the length of the infant's outstretched arms. The red and orange areas represent the infant's field of view, with the orange area being the center of the field of view. (b) Block diagram showing the process of learning hand regard.

To compare the predicted sensory feedback (corollary discharges) with the actual sensory feedback (visual inputs and proprioceptive inputs) and to distinguish the "hand" from the "other", a simplified "forward model" (Miall & Wolpert, 1996), which transforms the efference copy (output activities of the eight output units) into a corollary discharge, was incorporated into the previous model. Visual and proprioceptive feedback signals associated with the movements of the "hands" and corollary discharges became inputs for the input units during the next time step. The purpose of the previous study was to verify that the comparison between the predicted sensory feedback and the actual sensory feedback would make it possible to distinguish hands from other objects. By simplifying hand shape and hand movement control, the corollary discharges, visual inputs and proprioceptive inputs have been simplified, making their comparison easier.

A simple neural network was trained, using a modified real-time recurrent learning (RTRL) algorithm (Williams & Zipser, 1989), to manage time-varying inputs and outputs during hand regard. Let $z_i(t)$ denote the outputs of the $i$th unit in the network at time $t$; let $x_i(t)$ denote the outputs of the $i$th input unit in the network at time $t$; and let $y_i(t)$ denote the outputs of the $i$th hidden unit and the $i$th output unit in the network at time $t$, using the following equation:

$$z_i(t) = \begin{cases} x_i(t), & i \in I \\ y_i(t), & i \in H \cup O, \end{cases} \quad (1)$$

where $I$ represents an index set of input units, $H$ represents an index set of hidden units, and $O$ represents an index set of output units. The outputs from the hidden units and output units are updated according to the following equations:

$$y_i(t+1) = f_i(s_i(t+1)), \quad i \in H \cup O \quad (2)$$

$$s_i(t) = \sum_{j \in I \cup H \cup O} w_{ij} z_j(t), \quad (3)$$

where $w_{ij}$ represent the weights connecting the $i$th unit with the $j$th unit, $s_i(t)$ denotes the net input to the $i$th unit at time $t$, and $f_i$ is the sigmoid function:

$$f_i(s_i(t)) = 1/(1 + e^{-s_i(t)}). \quad (4)$$

The weight changes are calculated as follows:

$$\Delta w_{ij}(t) = \eta \sum_{k \in O} e_k(t) p_{ij}^k(t), \quad (5)$$

where $\eta$ is the learning rate, $e_k(t)$ represent motor-command errors, which are estimated as the coordinate value of each "hand" minus the coordinate value of the center position of the field of view, and $p_{ij}^k(t)$ is given by the following equation:

$$p_{ij}^k(t) = f_k'\big(s_k(t)\big)\Big(\delta_{ik} z_j(t-1) + \sum_{l \in H \cup O} w_{kl}\, p_{ij}^l(t-1)\Big). \tag{6}$$

Here, $\delta_{ij}$ denotes the Kronecker delta. In advance of using the RTRL algorithm, a set of input data and teaching signals must be prepared for the training phase, and input data must be prepared for the test phase at every time step; however, these requirements cannot be satisfied because the positions of the left and right "hands" and "other" objects change dynamically. Therefore, the motor-command errors (i.e., the differences between teaching signals and outputs) for each "hand" on the output units for every time step were estimated to be proportional to the coordinate value of each "hand" minus the coordinate value of the center position of the field of view, based on the method proposed by Kawato et al. (Kawato, Furukawa, & Suzuki, 1987). Because hand regard can be observed in blind infants (Freedman, 1964), the difference between the position of each "hand" and the center position of the field of view was computed by using the proprioceptively perceived position, instead of visual information. The weights in the network were updated every ten time steps during the training phase.

Little is known of what part of the brain is related to learning of hand regard and what kind of inputs and learning rule are used to perform that learning. In the previous study, to simulate the developments of hand regard, the RTRL algorithm was adopted. It is necessary to verify the effects of the selection of the algorithm and analysis conditions (input data, motor-command errors, update interval, etc.) on the learning of hand regard and the distinction between hands and others realized from it.

Figure 2a shows the visual attention (defined as "the state in which the infant's eyes are more than half open, their direction of gaze shifting within 30 seconds") of several subjects who were reared with virtually nothing aside from their own hands to view; accordingly, their visual attention could be interpreted as the frequency with which they view their own hands (White & Held, 1966). The visual attention of these infants increased sharply at approximately two months of age and was almost constant for the following six weeks. This result can be explained by the fact that sustained hand regard begins at approximately two months of age and continues during the same period;

therefore, infants spend considerable time watching their hands. A neural network was trained ten times, with randomly initialized weights in the range [-0.1, 0.1], using an RTRL algorithm, and the success rate, measured as the frequency with which the "hand" enters the center of the visual field during the training phase, was estimated. The ensemble average of the success rates, obtained after the training was performed ten times, is plotted in Fig. 2b. A comparison of the visual attention, plotted in Fig. 2a, and the success rate, plotted in Fig. 2b, shows that the trained model reproduced the sharp increase in success rates, which can be observed during the development of visual attention at approximately 60 days of age.

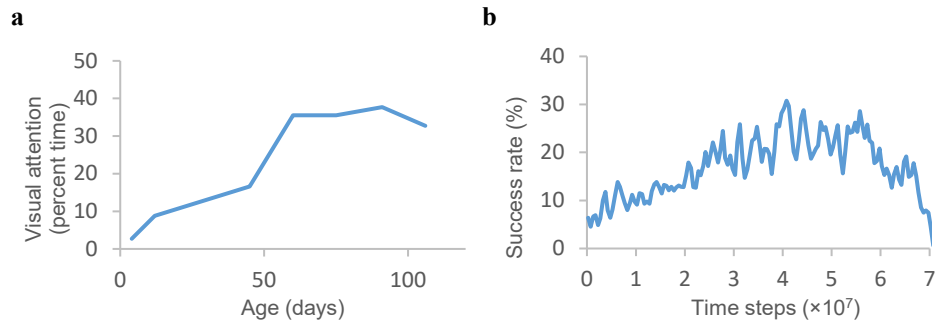

**Fig. 2.** Visual attention and success rate: (a) Development of visual attention for the subjects assigned to the control group. Each point represents "the average of two scores taken during successive two-week periods" (White & Held, 1966). (b) Plot of an ensemble average of success rates, obtained by training ten times.

The results of one trial, out of ten training trials, are shown in Fig. 3. A time series of the success rate (Fig. 3b) indicates repeated U-shaped development. Because the output activities of the hidden and output units are calculated using the sigmoid function, these output activities take values from 0 to 1. The color scale in Fig. 3a displays the range of these output activities. The colors of the squares in each panel of Fig. 3a show the output activities of the output units and hidden units, with the red or blue squares showing the output activity of output or hidden units, corresponding to values of 1.0 or 0.0, respectively.

The initial weights of the neural network were randomized in the range [-0.1, 0.1]. However, the hidden units were gradually interconnected with inhibitory weights. Most weights between the hidden units became inhibitory at $5.0\times10^3$ time steps (Fig. 3a); therefore, the output activities of the hidden units were close to zero. Then, the hidden units that excited each other appeared at $2.2\times10^6$ time steps, as shown by the red squares in Fig. 3a. This result is consistent with the definition of a cell assembly (i.e., a group of neurons that are strongly coupled by excitatory synapses) (Hebb, 1949). After the emergence of the cell assemblies, the configuration of cell assemblies changed each time U-shaped developments occurred (Fig. 3a). The output activities of the hidden units fluctuated significantly, with some inhibitory weights being transformed into excitatory ones, and the cell assembly appeared during the phase of U-shaped developments. These results show that the formation of cell assemblies is the local optimal solution for motor-command errors, which were estimated by the difference between the position of the "hand" and the center position of the field of view.

After the network was trained, whether "hand" and "other" could be distinguished was tested. A neural network was trained ten times, with weights initialized randomly. During the training phase for each of ten initializing weights, the network weights were saved every $1.0\times10^6$ time steps. The collection of success rates, which were calculated using the network weights that were saved every $1.0\times10^6$ time steps, resulted in a time series of success rates. A time series of success rates during the test phase was obtained ten times, by testing the network with the network weights that were saved every $1.0\times10^6$ time steps in response to the ten initializing weights. The test, which consisted of cases with various visual input values for "other" and the number of "others", was conducted. The "other" was arranged in the whole area during the training phase. In contrast, "others" were arranged and maintained in the field of view during the test phase; consequently, maintaining "others" in the field of view made it more difficult to distinguish between "hand" and "other".

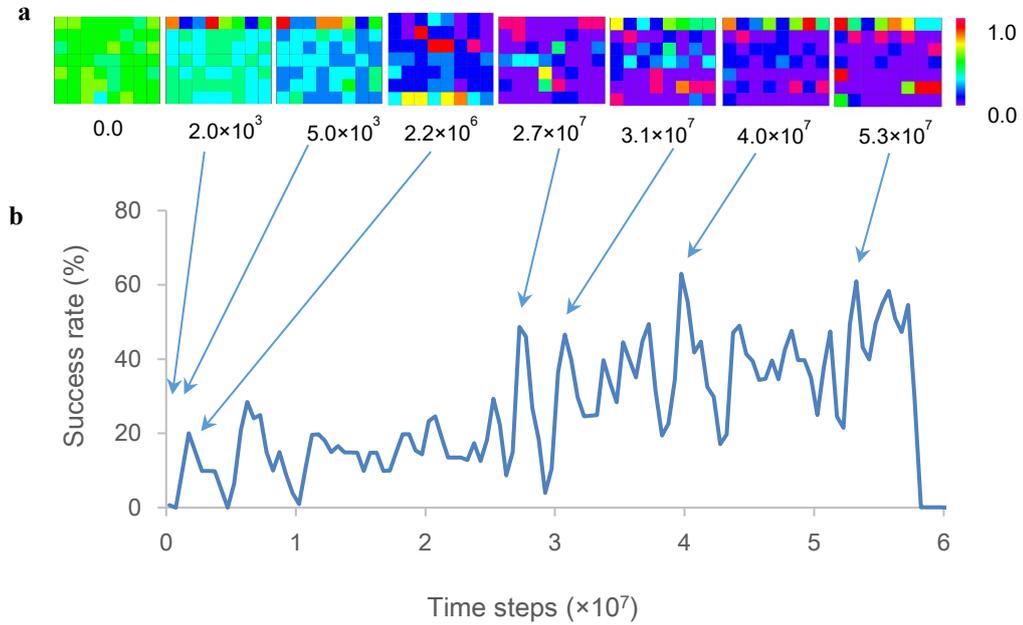

**Fig. 3.** Output activity and success rate: (a) The output activities resulting from one of ten training trials. Each panel represents the output activities of the hidden and output units at 0.0, $2.0 \times 10^3$, $5.0 \times 10^3$, $2.2 \times 10^6$, $2.7 \times 10^7$, $3.1 \times 10^7$, and $5.3 \times 10^7$ time steps. Squares in the top line, those in lines 2-4, and those in lines 5-7 of each panel show the output activities of the eight output units, the 24 hidden units related to the sense of self-ownership, and the 24 hidden units related to the sense of self-agency, respectively. (b) A representative time series of success rates, obtained by training ten times.

The ensemble averages of the success rates obtained by the ten testing trials for each case are plotted in Fig. 4a. Because case 1 represents the same conditions as those used for the training phase, the results for case 1 were similar to the results of the training phase (Fig. 2b). The visual input values of "other" in cases 1, 2, and 3 were equal to the visual input value of "other" in the training phase (i.e., 0.2), but the number of "others" were 1 (case 1), 5 (case 2), and 20 (case 3). In contrast, the visual input values of "other" in cases 4, 5, and 6 were equal to those of the right and left "hands" (i.e., 0.5), but the number of "others" were 1 (case 4), 5 (case 5), and 20 (case 6); that is, for case 1 and case 4, case 2 and case 5, case 3 and case 6, the visual input values of "other" had different values, but the number of "others" had the same value.

Figure 4a shows that the success rate decreased as the difficulty of the conditions increased, and whether the network could distinguish between "hands" and "others" was unclear. In particular, as can be seen in case 6, when there were as many as 20 "others" that were given the same visual input value as the "hand", identifying the "hand" took longer, even when the trained network had previously acquired the ability to distinguish between "hand" and "other". As a result, the success rate was low for case 6.

To show that the network acquired the ability to distinguish between "hands" and "other", the following test was conducted (Fig. 4b). As mentioned above, motor-command errors, which were estimated as the differences between the positions of the "hand" and the center position of the field of view, were computed using proprioceptively perceived positions. The weights were updated so that the "hand" could move to the field of view according to the proprioceptively perceived position of "hand" during the training phase; therefore, training could potentially be conducted using only proprioceptive inputs. The following tests were conducted to examine the effects of visual inputs and corollary discharges during training.

The condition for case 7, shown in Fig. 4b, was the same as for case 1, except that the visual input value of "hands" and the value of the corollary discharge were set to zero; therefore, case 7 represents a test that moved the "hand" into the field of view using only proprioceptive inputs, without visual inputs or corollary discharges. Similarly, the condition for case 8 was the same as for case 1, except that the visual input value of "hands" was set to zero, and the condition for case 9 was the same as for case 1, except that the value of corollary discharge was set to zero. The success rate was higher in the presence of both visual inputs and corollary discharges, indicating that the network acquired a new ability to increase the success rate by using visual inputs and corollary discharges. In other words, visual inputs, proprioceptive inputs, and corollary discharges were integrated, and the network acquired the ability to distinguish between the "hand" and "other". By distinguishing between the "hand" and "other", more efficient movement of the "hand" became possible.

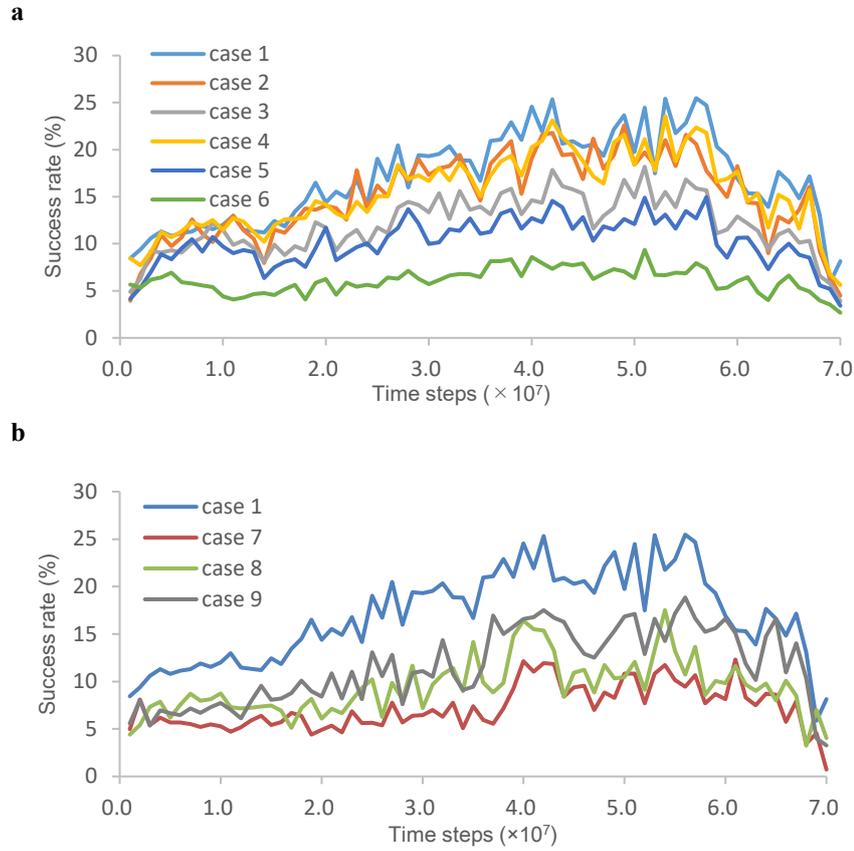

**Fig. 4.** Time series of success rates obtained by testing the network: (a) When "other" moved some squares in the field of view, the input unit corresponding to the square where "other" stayed received a visual input value. The visual input values of "other" and the number of "others" were 0.2 and 1 (case 1), 0.2 and 5 (case 2), 0.2 and 20 (case 3), 0.5 and 1 (case 4), 0.5 and 5 (case 5), and 0.5 and 20 (case 6). The visual input values of "other" in cases 1, 2, and 3 were equal to the visual input value of "other" in the training phase (i.e., 0.2). The visual input values of "other" in cases 4, 5, and 6 were equal to those of the right and left "hands" (i.e., 0.5). (b) Comparison of success rates for cases 1, 7, 8, and 9. The results of case 8 and case 9 were added to the figure from the previous study.

## 3. Methods

### 3.1 Temporal change in weights

The observed increases in the inhibitory weights between the hidden units indicate that the frequency at which weight changes take a negative value is greater than the frequency at which weight changes take a positive value. The weight changes were calculated from motor-command errors in the RTRL algorithm (Eq. (5)). Therefore, to elucidate the mechanism through which most of the weights between the hidden units in the previous model became inhibitory during the early stages of training, the relationship between the weight changes and the hand movements was examined. Because many different weights exist between hidden units, examining the relationship between the time change of each weight and hand movements was difficult. The weights in the network were updated every ten time steps during the training phase. To serve as an index that indicated the time change of the weight values, a difference obtained by subtracting the number of weight changes with negative values from the number of weight changes with positive values was adopted in the previous study (Takahiro Homma, 2019). This index was calculated every ten time steps from, training start to 600 steps, for the network whose training results are shown in Fig. 3. By examining the movements of both "hands" during this period and comparing the index with these movements, the causes for inhibitory weight generation were elucidated.

In this paper, the calculation of the index was extended from training start to 1,000 steps to examine changes in the index over longer periods. In addition, the temporal change, calculated by subtracting the number of weights (not weight change) with negative values from the number of weights with positive values during the entire training period, was also examined to show the relationship between the temporal change and the formation of cell assemblies.

## 3.2 Ablating the cell assemblies

In the present study, a group of hidden units with an output value of 0.9 or greater was defined as a cell assembly. These hidden units were strongly coupled by excitatory weights, which satisfied the definition of a cell assembly that was proposed by Hebb (Hebb, 1949). To evaluate the contribution of the information stored in the cell assemblies, the units belonging to the cell assemblies were ablated from the hidden units, as follows. Among the weights stored every $1.0 \times 10^6$ time steps in the previous study (Section 2), the weights for the hidden units with outputs of 0.9 or greater were set to zero. For the weights from any hidden unit in the cell assemblies to any input or hidden unit,

$$w_{ij} = 0, \; i \in I \cup H, \; j \in the\ set\ of\ all\ hidden\ units\ in\ cell\ assemblies, \quad (7)$$

and for the weights from any hidden or output unit to any hidden unit in the cell assemblies,

$$w_{ij} = 0, \; i \in the\ set\ of\ all\ hidden\ units\ in\ cell\ assemblies, \; j \in H \cup O. \quad (8)$$

Here, $I$ represents an index set of input units, $H$ represents an index set of hidden units, and $O$ represents an index set of output units. The ablation was conducted on the network weights that were saved every $1.0 \times 10^6$ time steps during the training phase, for each of ten initializing weights. As a result of this ablation, ten sets of weights were obtained, with the weights of the units belonging to the cell assemblies equal to zero.

## 3.3 Calculating the contribution ratio of the cell assembly

A test, which consisted of six cases with varying visual input values of "other" and varying numbers of "others", as shown in Fig. 4a, was conducted using the ten sets of weights that were obtained after ablating the cell assemblies. A time series of success rates during the test phase was obtained ten times by testing the network with the ten sets of weights for each case shown in Fig. 4a. Subtracting the time series of success rates obtained after ablating the cell assemblies from the time series of success rates obtained before ablating the cell assemblies (Fig. 4a, hereinafter referred to as success rate before ablation) resulted in a time series of success rates demonstrating the contributions of information stored in the cell assemblies (hereinafter referred to as the success rate contributed to by cell assemblies).

The contribution ratio of the cell assembly is defined as an index for measuring the degree to which the information stored in the cell assembly contributed to improvements in the success rate, and was calculated as follows. The time series of success rates consists of a collection of success rates calculated by the network weights that were saved every $1.0 \times 10^6$ time steps. Let the contribution ratio of the cell assembly equal the success rate contributed to by cell assemblies divided by the success rate before ablation:

contribution ratio of the cell assembly = success rate contributed by cell assemblies / success rate before ablation. (9)

The contribution ratio of the cell assembly was calculated using the success rates obtained by testing ten sets of weights for each of the six cases shown in Fig. 4, taking measurements every $1.0 \times 10^6$ time steps. The collection of contribution ratios for the cell assemblies, obtained every $1.0 \times 10^6$ time steps, resulted in a time series for the contribution ratio of the cell assembly. Consequently, the ensemble averages of the contribution ratio of the cell assembly were obtained using ten sets of weights for each case.

## 4. Results

### 4.1 Temporal change in weights

The relationship between weight changes and hand movements was examined during one trial, out of ten training phases, whose results are illustrated in Fig. 3. From the training start to 1,000 steps, the right "hand" reciprocated horizontally, and the left "hand" reciprocated between the upper right and lower left, across the field of view, so that both "hands" were within the field of view (Fig. 5).

Whether the weight changes between hidden units caused this reciprocation was investigated. The time series of the difference obtained by subtracting the number of weight changes with negative values from the number of weight changes with positive values (blue) between hidden units, every ten time steps, from training start to 1,000 steps, is plotted in Fig. 6. Moreover, the periods when the left "hand" (red) and the right "hand" (green) moved are also illustrated in Fig. 6. Although there is an exception for a portion of the movement period for the right "hand" (near 280 steps and 370 steps), Figure 6 shows a following relationship between the movements of the "hands" and this difference in weight changes.

As the weights between the hidden units were updated by positive weight changes, the values of the weights between hidden units increased; consequently, the outputs from the hidden layer to the output layer also increased until the outputs of the output units exceeded the threshold of the outputs required for the "hands" to move, causing the "hands" to move toward the opposite points, across the field of view. Figure 6 indicates that the first movement of both "hands" occurred at approximately 30 time steps. Both "hands" passed the field of view at approximately 40 time steps. The motor-command errors, $e_k(t)$ in Eq. (5), for each "hand" on the output units were estimated to be proportional to the coordinate values of each "hand" minus the coordinate value of center position of the field of view (Section 2). When both "hands" passed the field of view, the coordinate values of both "hands" and the center position of the field of view became equal; therefore, the motor-command error $e_k(t)$ became zero and $\Delta w_{ij}(t)$, in Eq. (5), also became zero at 40 time steps, as shown in Fig. 6. When the

"hands" moved toward the opposite points, across the field of view, at approximately 50 time steps, the signs of motor-command error $e_k(t)$ for each "hand" were reversed, and $\Delta w_{ij}(t)$ became a negative value, as shown in Fig. 6. Therefore, the weights between hidden units decreased, and the outputs of hidden units also decreased. Consequently, the outputs of the output units fell below the threshold of outputs required for the "hands" to move, and the "hands" stopped, without moving further, at approximately 50 time steps. The difference obtained by subtracting the number of weight changes with negative values from the number of weight changes with positive values decreased until 70 time steps and then began to increase. The weights between hidden units were updated by the positive weight changes, and the "hands" moved again, in the opposite direction across the field of view, at approximately 110 time steps. By repeating this process, reciprocating motion of the "hands" across the field of view occurred.

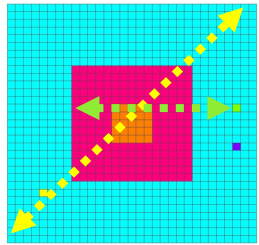

**Fig. 5.** Movements of both hands from training start to 1,000 steps.

Figure 6 shows that the frequency at which the weight changes became negative was higher than the frequency at which the weight changes became positive. This difference in frequency resulted in most of the weights between the hidden units representing inhibitory inputs during the early stages of training; inhibitory weights appeared to be generated by trial and error attempts to maintain both "hands" into the field of view.

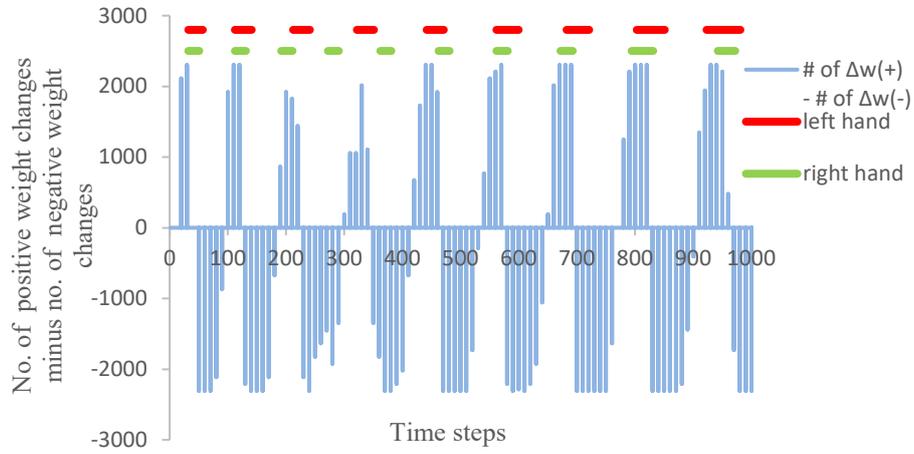

**Fig. 6.** Time series of the number of positive weight changes minus the number of negative weight changes and the movement periods for both "hands".

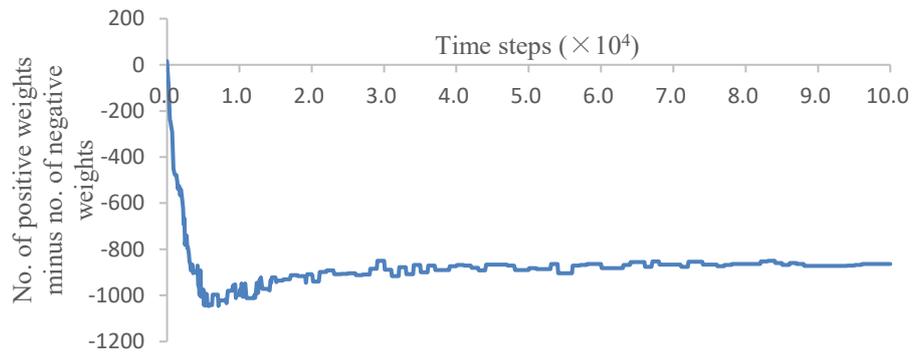

**Fig. 7.** Time series of the number of positive weights minus the number of negative weights.

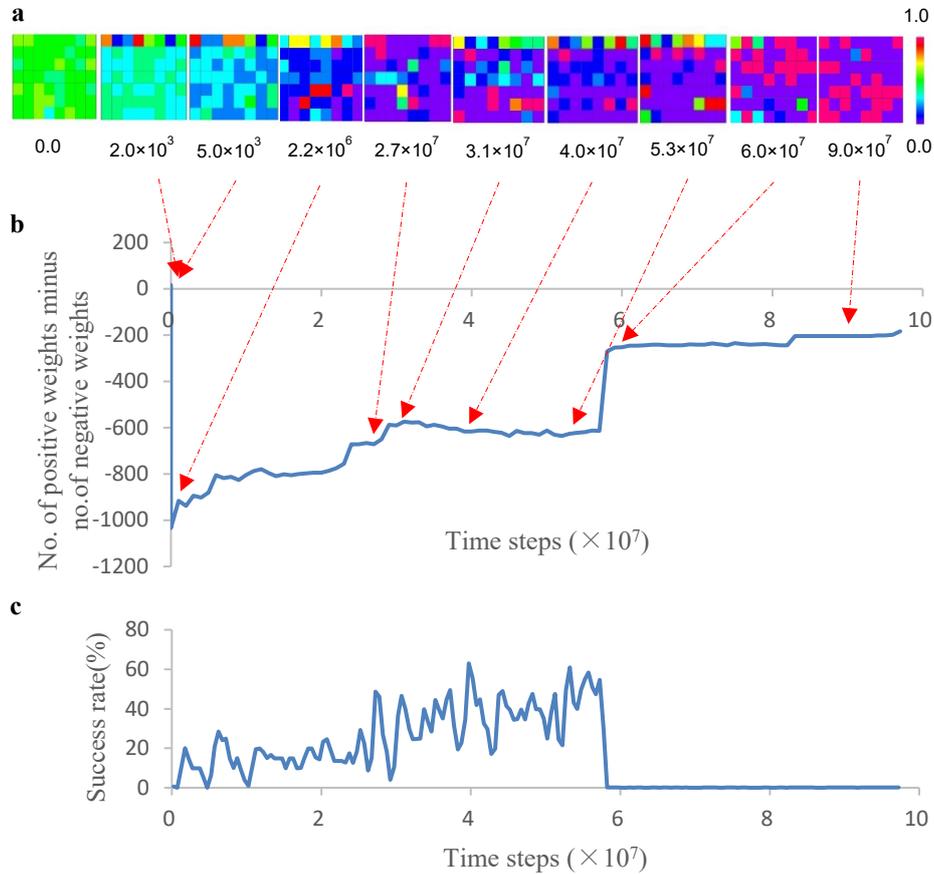

**Fig. 8.** Output activity, a time series of the number of positive weights minus the number of negative weights, and the success rate through 1.0×10⁸ time steps: (a) Output activities resulting from one of ten training trials. Each panel represents the output activities of hidden and output units, at 0.0, 2.0×10³, 5.0×10³, 2.2×10⁶, 2.7×10⁷, 3.1×10⁷, 5.3×10⁷, 6.0×10⁷, and 9.0×10⁷ time steps. Squares in the top line, those in lines 2–4, and those in lines 5–7 of each panel represent the output activities of the eight output units, the 24 hidden units related to a sense of self-ownership, and the 24 hidden units related to a sense of self-agency, respectively. (b) Time series of the number of positive weights minus the number of negative weights. (c) One of the time series for success rates, obtained from ten training trials.

The decrease in the values of weights continued until the "hands" were able to move into the field of view. Figure 7 shows a time series of the number of positive weights (not weight changes) minus the number of negative weights between the hidden units. As depicted in this figure, the minimum value of this number became -1,046 at 5,700 time steps. Because the total number of weights between the hidden units was 1,152, most of the weight values were negative at 5,700 time steps. The "hand" began to enter the field of view at approximately 4,000 steps, and after 6,000 steps, when the number of positive weights increased, the "hand" entered the field of view without trial error.

The results obtained by further extending the time series for the number of positive weights minus the number of negative weights to $1.0 \times 10^8$ time steps are shown in Fig. 8b. The number of positive weights increased until $2.2 \times 10^6$ time steps, at which point the cell assembly (red squares) appeared, and then the positive weights continued to increase until approximately $3.0 \times 10^7$ time steps. Thereafter, the number of positive weights was almost constant but increased rapidly at $5.8 \times 10^7$ time steps. Along with this increase, the number of units belonging to the cell assembly also rapidly increased (Fig. 8a). The values of the outputs from many of the hidden units belonging to the cell assembly approached 1.0 for any inputs. As a result, the "hand" could not enter the field of view, which indicates that synaptic transmission was saturated. The trajectories of both "hands" during 100-time-step periods at $6.0 \times 10^7$ time steps and at $9.0 \times 10^7$ time steps (Fig. 9) showed that both "hands" moved upward and could not stay in the field of view; consequently, the success rate became almost zero.

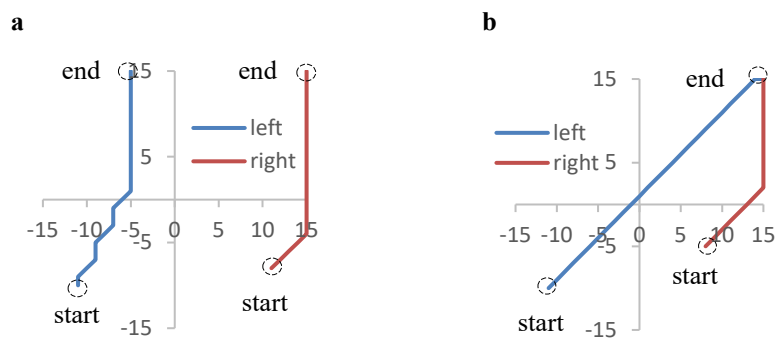

**Fig. 9.** Trajectories of both "hands" during a 100-time-step period: (a) $6.0 \times 10^7$ time steps. (b) $9.0 \times 10^7$ time steps.

*4.2 Contribution of the cell assembly*

During the training phase, hidden units were gradually interconnected with inhibitory weights until the cell assembly appeared at $2.2\times10^6$ time steps (Fig. 3 and Fig. 8). Figure 10 shows the distribution of weight values at the initial step, and at $2.1\times10^6$ time steps immediately before the cell assembly appeared. Since the weights were initialized randomly, the number of positive initial weights and the number of negative ones were almost the same in the range [-0.1, 0.1]. In contrast, at $2.1\times10^6$ time steps, most of the weights become inhibitory and those absolute values become larger than the initial values. The cell assembly is formed according to the principle of the soft Winner-Take-All or k-Winner-Take-All (Maass, 2000; Majani, et al., 1989) only after most of the weights become inhibitory in this way.

A group of hidden units with an output value of 0.9 or greater was defined as a cell assembly in this paper. Figure 11 shows the values of the weights between two units that belong to cell assemblies, from one trial, out of ten training phases, for which the results are illustrated in Fig. 3. These values are plotted every $1.0\times10^6$ time steps in Fig. 11. Therefore, although the cell assembly emerged at $2.2\times10^6$ time steps, these values are plotted starting from $3.0\times10^6$ time steps. The weights for a neural network were initialized randomly in the range [-0.1, 0.1]. The hidden units that belong to a cell assembly and the values of the weights between those units changed during the training phase. While some weights had negative values, most weights had large positive values, which satisfied the definition of a cell assembly (i.e., a group of neurons that are strongly coupled by excitatory synapses) that was proposed by Hebb (Hebb, 1949).

In the above training trail, a cell assembly composed of four hidden units first appeared at $2.2\times10^6$ time steps (Fig. 3 and Fig. 8). The realization of a soft Winner-Take-All phenomenon is represented by the changes in the values of the weights between these four hidden units (Fig. 12). After the weights between the hidden units were randomly initialized in the range [-0.1, 0.1], most of the weights between the hidden units had negative values (Fig. 7 and Fig. 8). However, Figure 12 shows that only the values of the weights between these four hidden units increased and that these values increased sharply at $2.2\times10^6$ time steps, which coincides with the initial appearance of the cell assembly. For the first unit, the weight of the self-loop (i.e., 1→1, in Fig. 12a and b) increased first. However, the weights of the other units gradually strengthened each other (Fig. 12c-h). These units became multiple winners of an excitatory and inhibitory competition. After $2.2\times10^6$ time steps, the arrangement of the cell assemblies changed with training, as shown in Fig. 3 and Fig. 8. As the arrangements of the cell assemblies changed, it is likely that the information stored within the cell assembly also changed. The content of the stored information was estimated from the contribution ratio of the cell assembly, as follows.

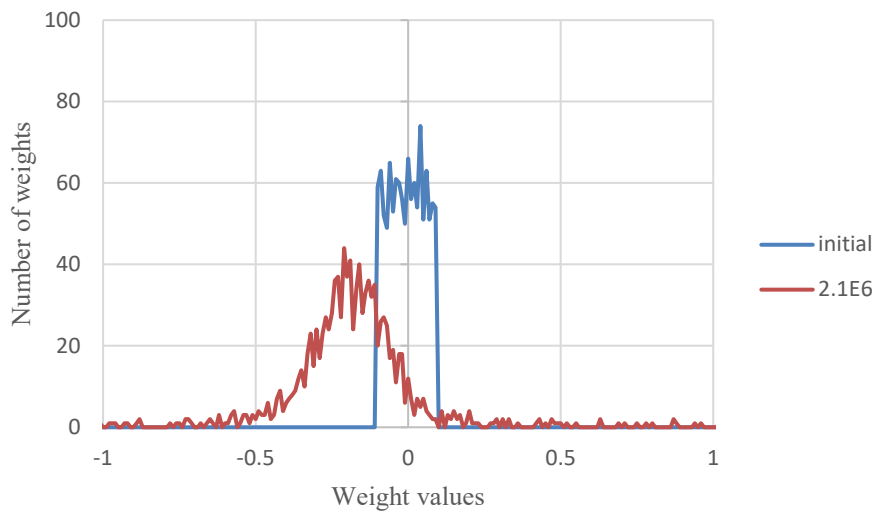

**Fig. 10.** Distribution of weight values at the initial step and at $2.1\times10^6$ time steps.

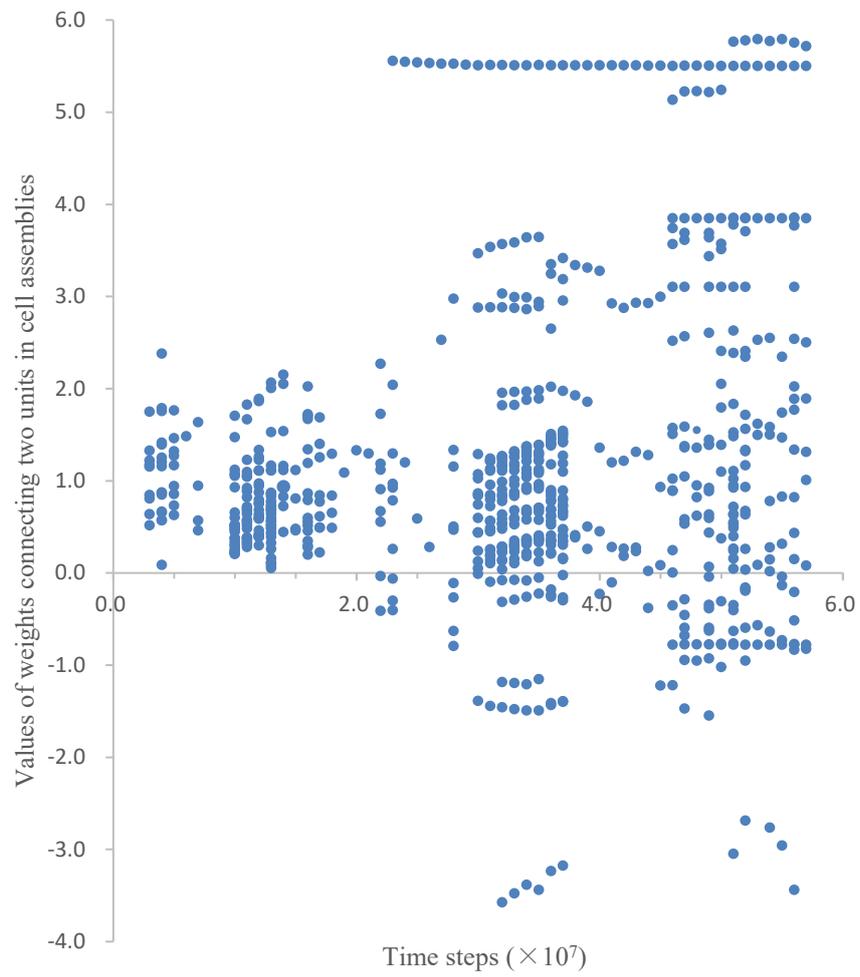

**Fig. 11.** Values of the weights between two units that belong to cell assemblies. The average of all weight values was 1.2.

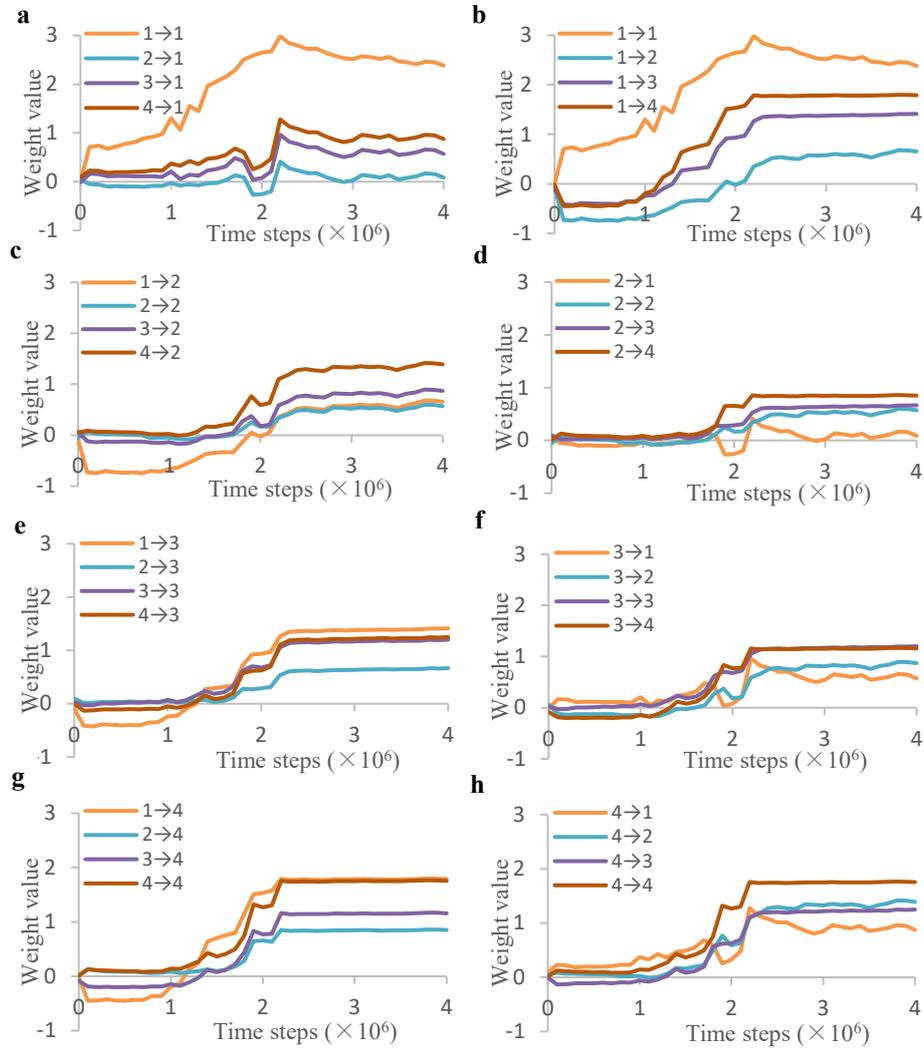

**Fig. 12.** Time series of weight values between two units that belong to cell assemblies: Time series of weight values for the following connections: (a) from four units to the first unit; (b) from the first unit to four units; (c) from four units to the second unit; (d) from the second unit to four units; (e) from four units to the third unit; (f) from the third unit to four units; (g) from four units to the fourth unit; and (h) from the fourth unit to four units. To more easily grasp the changes in the weights connected to each unit, duplicate results are presented in each panel.

Figure 13 shows the ensemble averages for the contribution ratios of the cell assembly, obtained using ten sets of weights for each of the six cases in Fig. 4. This figure shows that the values of the contribution ratios of cell assembly varied among the six cases until $4.0\times10^7$ time steps. In contrast, after $4.0\times10^7$ time steps, the values of the contribution ratios of cell assembly became almost constant, and the variation between cases was small. The average contribution ratio from case 1 to case 6, after $4.0\times10^7$ time steps, was 0.82. The small variation in the contribution ratios of cell assembly after $4.0\times10^7$ time steps can also be observed in Fig. 14, which shows the temporal changes in the standard deviation of these values among the six cases. The input data for the neural network included visual inputs for the "hands", proprioceptive inputs, corollary discharges, and visual inputs for "others". Among the six cases shown in Fig. 4, only visual inputs of "others" (the visual input value of "other" and the number of "others") were changed. Therefore, the small variations in the contribution ratios of cell assembly among these six cases after $4.0\times10^7$ time steps indicate that the contribution ratio was not dependent on the visual input for "others".

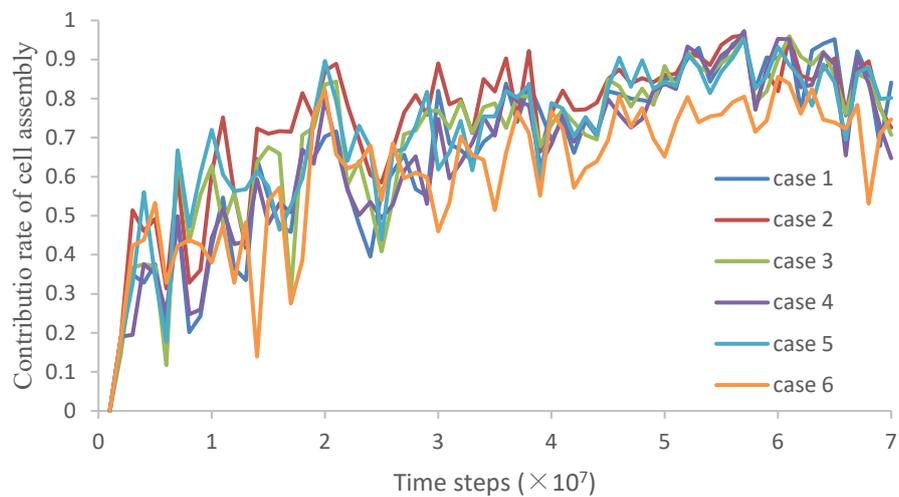

**Fig. 13.** Time series of contribution ratios of cell assembly.

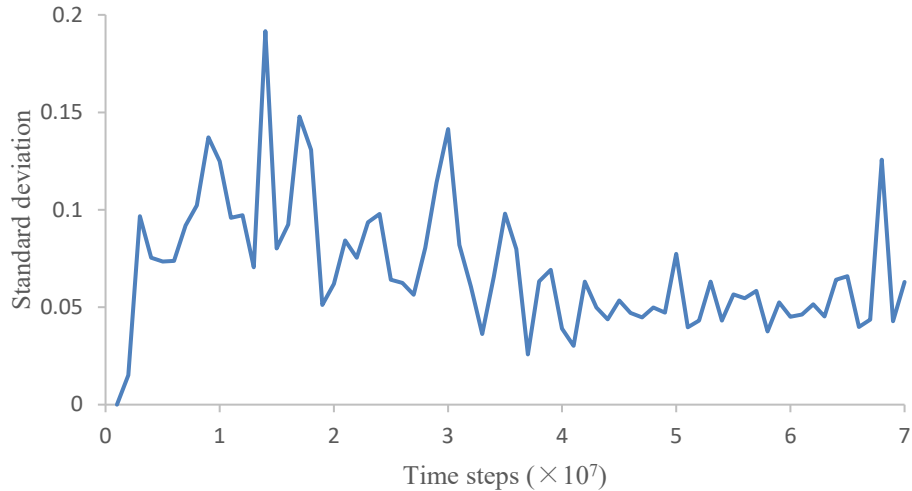

**Fig. 14.** Temporal changes in the standard deviations for the contribution ratios of cell assembly among six cases.

## 5. Discussion and conclusion

In the present paper, the mechanism underlying inhibitory weight generation that plays an important role in the formation of cell assemblies, during the learning of hand regard, was determined. Through trial and error attempts to maintain both "hands" within the field of view, the frequency at which weight changes became negative was higher than the frequency at which weight changes became positive; consequently, inhibitory weights were generated (Fig. 6). Infants tend to move one hand into their field of view, rather than both hands, due to the effects of the asymmetrical tonic neck reflex (ATNR). Both "hands" moved into the field of view in this model because the ATNR was not incorporated into the present model. Further refinement of the model, incorporating ATNR, will occur in a future study.

Decreases in the values of weights continued until the "hands" were able to move into the field of view without trial error. Then, the hidden units that composed the cell assembly represented multiple winners of an excitatory and inhibitory competition, which could be understood as a soft Winner-Take-All or k-Winner-Take-All phenomenon (Fig. 8 and Fig. 12).

In the RTRL algorithm used for the learning of hand regard, the overall network error, calculated from the motor-command errors $e_k(t)$ in Section 2, was minimized (Williams & Zipser, 1989). Neurons can experience long-term depression (LTD), which prevents the saturation of synaptic transmission by downregulating synaptic functions (Kandel, Schwartz, Jessell, Siegelbaum, & Hudspeth, 2013). However, because this algorithm does not include an LTD function, it reached a local minimum where synaptic transmission was saturated. Consequently, the "hands" did not move into the field of view and hand regard disappeared (Fig. 8 and Fig. 9). However, hand regard also disappears at approximately four months in infants. A future study will clarify whether the disappearance of hand regard occurs due to the saturation of synaptic transmission or to genetic information.

To evaluate the contributions made to hand movement by the information stored in the cell assemblies, the units belonging to the cell assemblies were ablated from the hidden units. Figure 13 shows that the values of the contribution ratios of cell assembly became almost constant, at an average value of 0.82, after $4.0 \times 10^7$ time steps. This result indicates that in all six cases, the "hand" has moved into the field of view primarily due to the output of the cell assembly. Figure 13 also illustrates that the contribution ratio of cell assembly was not dependent on the visual inputs for "others" after $4.0 \times 10^7$ time steps. The input data for the neural network included visual inputs for the "hands", proprioceptive inputs, corollary discharges, and visual inputs for "others"; therefore, the contribution ratio of cell assembly primarily depended on the visual inputs of the "hands", proprioceptive inputs, and corollary discharges. Inputs for the "hand" were selected from among multiple inputs to the units belonging to the cell assembly (i.e., the "hand" was distinguished from the "others"), and the outputs of these units moved the "hand" to the center of the field of view. This result is consistent with the results of our previous study, which showed that visual inputs, proprioceptive inputs, and corollary discharges were integrated and that the network acquired the ability to distinguish between the "hand" and "other" (Section 2). Therefore, after $4.0 \times 10^7$ time steps, the network was considered to have acquired the ability to distinguish the "hand" from "other", and the information regarding this ability was stored in the cell assembly. In case 6, twenty "others", with the same visual input values as the "hand", were arranged and maintained in the field of view during the test phase. Identifying the "hand" among all "others" requires identifying an object that matches the visual inputs, proprioceptive inputs, and corollary discharges. Under the conditions for case 6, identifying the "hand" took a longer time, and the output of the units belonging to cell assemblies was not able to move the "hand" into the field of view. As a result, the success rate (Fig. 4a) and the contribution ratio of cell assembly (Fig. 13) declined for case 6.

Comparisons between the observed results and the simulation results shown in Fig. 2 suggest that the period after $4.0 \times 10^7$ time steps corresponds with the period during which sustained hand regard occurs. During the period of sustained hand regard, infants may acquire the ability to move their hand into the field of view after distinguishing their hand from others, using the information stored in the cell assembly.

Because a non-invasive observation method for cell assemblies has not yet been established, observing the generation of cell assemblies in infant's brains during hand regard or comparing the biological process with the present results is not possible. Procedural learning consists of both motor and nonmotor learning. Motor movements are first planned, based on the geometry of the environment, and the control of motor movements depends on feedback from the visual and proprioceptive systems (Knowlton, Siegel, & Moody, 2017). Motor learning based on the geometry of the environment may involve trial and error, similar to the learning of hand regard in the present model. Therefore, if and when a non-invasive observation method for the generation of cell assemblies is established in the future, comparisons between the observed cell assembly process during the learning of hand regard and the results of this paper may become possible; furthermore, the proposed method used in this paper may be applicable to studies of cell assemblies during other types of motor learning.

## Acknowledgements

The author thanks Yutaka Nakama for permitting the use of his visualization program (NAK-Post). This research did not receive any specific grant from funding agencies in the public, commercial, or not-for-profit sectors.